\documentclass[twocolumn,showpacs,amsmath,amssymb,prl,superscriptaddress]{revtex4}
\usepackage{graphicx}
\usepackage{color}

\begin{document}

\title{Many-body effects in TiSe$_2$: can GW describe an excitonic insulator?}
%\title{Many-body effects in TiSe$_2$: indications for an excitonic insulator from GW}
\author{M. Cazzaniga}
\affiliation{Universit\`a degli Studi di Milano, Physics Department, Milano, Italy}
\affiliation{European Theoretical Spectroscopy Facility (ETSF)}
\author{H. Cercellier}
\affiliation{Institut N\'eel, CNRS and UJF, Grenoble, France}
\author{M. Holzmann}
\affiliation{LPTMC, CNRS and UPMC, Jussieu and LPMMC, CNRS and UJF, Grenoble, France}
\affiliation{European Theoretical Spectroscopy Facility (ETSF)}
\author{C. Monney}
%\affiliation{Research Department Synchrotron Radiation and Nanotechnology, Paul Scherrer Institut, CH-5232 Villigen PSI, Switzerland}
%\affiliation{Res. Dep. Synchr. Rad. and Nanotechn., Paul Scherrer Institut, Villigen PSI, Switzerland}
\affiliation{Synchrotron Radiation and Nanotechnology Department, Paul Scherrer Institut, Villigen, Switzerland}
\author{P. Aebi}
%\affiliation{D\'epartement de Physique and Fribourg Center for Nanomaterials, Universit\'e de Fribourg, CH-1700 Fribourg, Switzerland}
\affiliation{D\'epartement de Physique and Fribourg Center for Nanomaterials, Universit\'e de Fribourg, Fribourg, Switzerland}
\author{G. Onida}
\affiliation{Universit\`a degli Studi di Milano, Physics Department, Milano, Italy}
\affiliation{European Theoretical Spectroscopy Facility (ETSF)}
\author{V. Olevano}
\affiliation{Institut N\'eel, CNRS and UJF, Grenoble, France}
\affiliation{European Theoretical Spectroscopy Facility (ETSF)}

\date{\today}

\begin{abstract}
We present both theoretical \textit{ab initio} GW and experimental angle-resolved photoemission (ARPES) and scanning tunneling (STS) spectroscopy results on TiSe$_2$.
With respect to the density-functional Kohn-Sham metallic picture, the many-body GW self-energy leads to a $\approx 0.2$ eV band gap insulator consistent with our STS spectra at 5 K.
The band shape is strongly renormalized, with the top-of-valence moved towards a circle of points away from $\Gamma$, arising in a \textit{mexican hat} feature typical of an  \textit{excitonic insulator}.
Our calculations are in good agreement with experiment.
%our ARPES and STS measurements, as well as  previous EELS experiments.
\end{abstract}

%71.10.-w

\maketitle

\paragraph{Introduction} 
Crystalline solids are classified as either insulators  or metals, 
depending on the existence or absence of a finite energy gap in the electronic excitation spectrum.
Within the independent particle picture, metallic or 
insulating behavior is explained in terms of partially or completely filled valence bands, but
electronic  correlations may overcome this ordinary paradigm and lead to
new mechanisms for  transforming a metal into an insulator or \textit{viceversa}.
From the early works of Mott, Kohn, and Hubbard \cite{Mott,Kohn1,Hubbard}, 
several systems showing such exotic behavior have been supposed to exist in nature.
One of them is the
so-called \textit{excitonic insulator} \cite{Mott,KeldyshKopaev,Kohn,Jerome,HalperinRice}, first predicted and studied in the works of Mott, Kopaev, Keldysh and Kohn and not to be confused with the more celebrated Mott-Hubbard insulator.
Here we study the possibility of an excitonic insulator in titanium diselenide (TiSe$_2$) \cite{Wilson}, focusing on the question whether \textit{ab initio} many-body GW theory \cite{Hedin,StrinatiMattauschHanke} can describe photoemission spectra of such an excitonic phase.

At a non-interacting one-electron level, an excitonic insulator would present a semimetallic or semiconducting band structure with sufficiently small (typically few tens of meV) overlap or gap, and  reduced number of free carriers.
Consequently, the Coulomb interaction between particles is only weakly screened.
Electrons and holes can then spontaneously bind into non-conducting excitons and form a 
new ground state of lower energy than the normal phase.
Exciton condensation may also lead to the formation of charge-density waves (CDW) of purely electronic origin.
On the semimetal side, model descriptions of the system %transition
are based on a BCS-like approach
where  the role of electron-electron Cooper pairs is
 taken by electron-hole excitons.
 Similar to BCS superconductors, an energy gap opens of the same order of magnitude as the binding energy of the pair
 \cite{Zittartz}
(depending on the underlying band structure, e.g. 
effective mass anisotropy and multivalley effects).

So far there is no clear-cut experimental proof of the excitonic insulator existence in nature.
Nevertheless, several experimental obervations point to TiSe$_2$ as one of the most probable candidates \cite{DiSalvo,Pillo,Kidd,Herve1,MonneyPLD,Herve2}.
Indeed a CDW has been observed in the ground state of TiSe$_2$, like in other transition-metal dichalcogenides,
but, so far, there is no consensus on 
 the interpretation of the ground state and its CDW. In particular, the exact role played by excitons is still controversial. 
Either a Jahn-Teller effect \cite{Hughes,Rossnagel} or a correlation mechanism leading to the excitonic insulator \cite{Pillo,Kidd} have been invoked.
Recently, it was suggested that the ground state and the resulting periodic lattice distortion (PLD) that occurs below 202 K are 
primary due to electron-phonon coupling, which is only enhanced by the presence of incoherent excitons \cite{vanWezel}.
On the other hand, photoemission spectra are consistent with the excitonic insulator scenario  \cite{Herve1},
and the interaction between an exciton condensate and phonons can reproduce the observed atomic displacements in the PLD with good agreement \cite{MonneyPLD}.

In this work we 
provide  theoretical evidence from \textit{ab initio} many-body GW calculations
supporting the excitonic insulator scenario in TiSe$_2$ and verify 
our calculations against angle-resolved photoemission spectroscopy (ARPES) measurements and scanning tunneling spectra (STS).
STS at 5 K clearly show a bandgap of $\simeq 0.15$ eV and point to TiSe$_2$ as a semiconductor.
With respect to the one-electron mean-field density-functional theory (DFT) band structure, where TiSe$_2$ is a metal, the GW many-body self-energy opens a band gap of $\simeq 0.2$ eV, thus leading to an insulator.
Furthermore, the band shape  at high symmetry points is strongly renormalized.
In particular, the top-of-valence is moved from $\Gamma$ toward a circle of points away from $\Gamma$,
building up a \textit{mexican hat} feature characteristic of an excitonic insulator, as predicted by Kohn \cite{Kohn}.
ARPES spectra indicate a flattening of the topmost valence bands before $\Gamma$, compatible with the mexican hat picture.
The GW bands we obtain are in good agreement with ARPES measurements, and also our STS spectra compare more favorably with the  GW density-of-states (DOS) than with DFT.
The calculated dielectric function describing the screening properties of the system, is in good agreement with previously measured energy-loss spectra (EELS) \cite{BuslapsJohnson}.
The overall picture arising from our results indicates a possible ground state instability in TiSe$_2$ of {\em electronic}  origin.
Further, our calculations show that bare GW is able to reproduce the photoemission band structure of systems with strong excitonic renormalizations, although, in general, vertex corrections beyond GW are necessary to properly describe excitons in optical spectra \cite{bse}.

\begin{figure}
 \includegraphics[width=\linewidth]{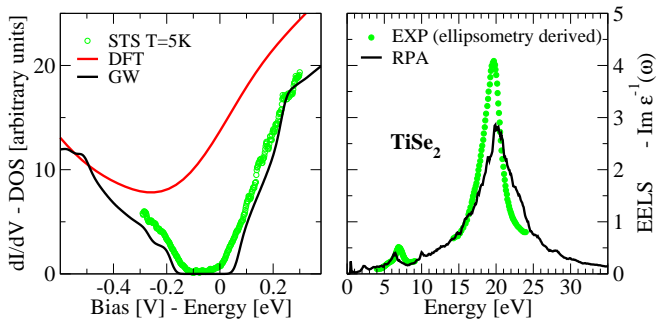}
 \caption{Left panel: STS $dI/dV$ experimental spectra (green circles) vs GW (shifted to account for finite impurity doping of the sample, black curve) and DFT (zeroed at $E_F$, red curve) DOS for TiSe$_2$.
Right panel: Energy-loss function. Green dots: experiment \cite{BuslapsJohnson} (derived from ellipsometry); black curve: present theoretical RPA result.}
 \label{elf+sts}
\end{figure}

\paragraph{Theory}
Our first step is a standard, numerically well converged \cite{calculation-details} ground-state DFT-LDA calculation of the total energy, the electronic density, and the lattice parameters within the non-reconstructed primitive-hexagonal crystal structure.
The DFT calculation provides the Kohn-Sham (KS) electronic structure that we use for  calculating the Green's function $G$ and the dynamically screened interaction $W(\omega)=\varepsilon^{-1}(\omega) v$ --- defined as  the bare Coulomb interaction $v$ screened by the dynamical dielectric function $\varepsilon^{-1}(\omega)$ ---, both entering into the self-energy in the GW approximation \cite{Hedin,StrinatiMattauschHanke},
\[
  \label{GWselfenergyomega}
  \Sigma^{\rm GW}(r,r',\omega) = \frac{i}{2 \pi}
  \int_{-\infty}^{\infty} d\omega' \,
  G(r,r',\omega-\omega')
  W(r,r',\omega')
  .
\]
Within a perturbative approach to first-order in $\Sigma-v_{xc}^{\rm LDA}$,
we then calculate the GW quasiparticle energies, expanding $\Sigma$ around the KS energies $\epsilon^{\rm KS}_{nk}$, 
taken as a zero-th order approximation for the true quasiparticle energies,
\[
  \epsilon^{\rm GW}_{nk} = \epsilon^{\rm KS}_{nk} +
    Z \big\langle nk \big| \Sigma^{\rm GW}\big(\omega=\epsilon^{\rm KS}_{nk} \big) - v_{xc}^{\rm LDA} \big| nk \big\rangle
  ,
\]
where $Z = (1 - \left. \partial \Sigma^{\rm GW} / \partial \omega \right|_{\omega=\epsilon^{\rm KS}_{nk}} )^{-1}$.

In the standard GW approach, the dielectric function $\varepsilon^{-1}(\omega)$ is calculated in the RPA approximation.
Since screening is of crucial importance for exciton formation, we have checked the RPA energy-loss function $-\Im \varepsilon^{-1}(\omega)$ against  the results of ellipsometry experiments \cite{BuslapsJohnson} (see Fig.~\ref{elf+sts}).
From the good agreement with experiment, we are confident that the used RPA screening accurately describes the 
physical situation.  Notice that the effective screening of our \textit{ab initio} RPA calculation is much weaker
compared to the Lindhard function, used in the model calculation of Ref.~\cite{Herve2},
so that we expect that exciton formation is even more favored.

\paragraph{Experiment}
Photon-energy dependent ARPES measurements were carried out at the SIS beamline of the Swiss Light Source synchrotron, using a Scienta SES-2002 spectrometer with an overall energy resolution better than 10 meV, and an angular resolution better than 0.5 $\deg$.
The data shown here have been collected at 65 K on two samples from the same batch with a slight Ti overdoping, resulting in a small population of the electron pocket at the L point and a consequent shift of the valence band.
%The  data of Fig.~\ref{zoom} correspond to the intensity maps of Ref.~\cite{Herve1}. %% arpes-eliminated
Scanning tunneling spectroscopy (STS) spectra were measured at low temperature by use of a OMICRON LT-STM with a lock-in technique.
The estimated energy resolution is about 10 meV.

%% arpes-eliminated
%\begin{figure}
% \includegraphics[width=\linewidth]{gwarpes}
% \caption{Left panel: zoom around $\Gamma$ of the second derivative of the ARPES signal with respect to $\omega$ at $T=65$ (left) and 251 K (right panel) vs GW bands (lines). GW bands have been down shifted to account for electron overdoping.
%}
% \label{zoom}
%\end{figure}

\begin{figure*}
 \includegraphics[width=\textwidth]{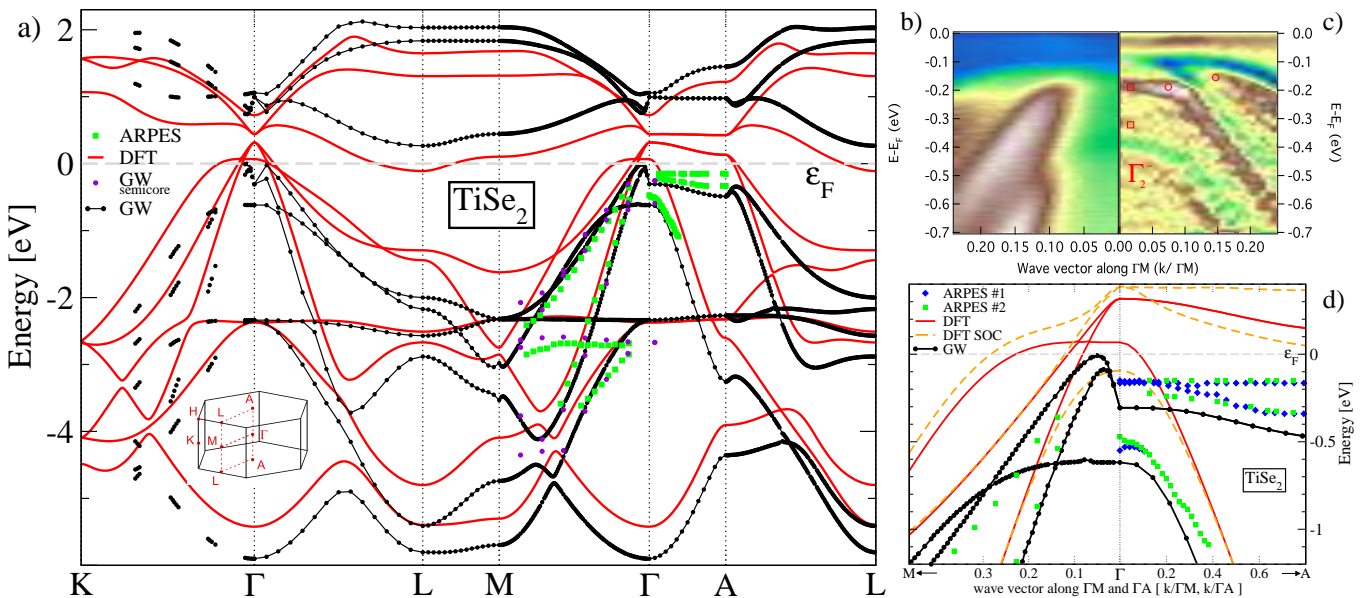}
 \caption{TiSe$_2$ band plot (\textit{a} panel) and its zoom (\textit{d}) at $\Gamma$ (M-$\Gamma$-A directions). Red lines: DFT-LDA KS electronic structure; yellow dashed lines: DFT plus spin-orbit correction (SOC); black dots and lines: GW band plot; violet dots: GW results including Ti 3$s$3$p$ electrons in valence; blue diamonds and green squares: sample 1 and 2 ARPES experimental data at $T = 65$ K. Panel \textit{b}: ARPES signal $I(k,E)$; and \textit{c}: its derivative $d^2I/dE^2$ taken at $T=65$ K around $\Gamma$, \textit{both} in the $\Gamma$-M direction.
}
 \label{bandplot}
\end{figure*}

\paragraph{Results: STS and GW band gap}
The results of our STS measurements at 5 K are given in Fig.~\ref{elf+sts}.
We plot the differential conductance $(dI/dV)$ as a function of the bias $V$. 
For small enough biases and slowly varying tunneling matrix elements, the differential conductance gives an image of the local density-of-states (DOS) that can be compared with calculations.
At small bias the experimental $dI/dV$ vanishes, indicating a gap of $\simeq 0.15$ eV.
With respect to the long debated TiSe$_2$ semimetal-semiconductor question \cite{Rasch}, our STS spectra, the first to date to our knowledge, provide a \textit{clear indication of semiconducting character}, at least at low temperature.
The STS gap is in quantitative agreement with recent ARPES experiments \cite{Rossnagel,Pillo,Kidd,Qian,Rasch}.
In Fig.~\ref{elf+sts} one can see that the calculated DFT DOS never vanishes, thus indicating a metal.
On the other hand, the GW DOS presents a well defined gap, like in STS, with just only a slight overestimation ($\simeq 0.2$ eV).
Contrary to DFT, GW found TiSe$_2$ a very small gap insulator, in agreement with the experiment.
The left-valence and right-conduction DOS profiles are also well reproduced by GW.
%A step at -0.3 eV, associated to the onset of the second valence band in the mexican hat, and a shoulder at 0.25 eV associated to a flattening of the conduction band at M, seem to be in correspondence with features in the experiment, another indication in favor of the GW mexican hat scenario.

\paragraph{Results: the band plot}
In Fig.~\ref{bandplot}\textit{a} we report the band plot of TiSe$_2$ based on DFT-LDA and GW calculations, together with experimental ARPES data.
According to the DFT KS electronic structure, TiSe$_2$ is metallic with a band overlap of 0.4 eV between the top of the pocket of holes at $\Gamma$ and the bottom of the pocket of electrons at L.
The GW electronic structure then leads to a general shift of the bands, negative for the lowest and positive for the highest, and opens a band gap of around 0.2 eV.
Therefore, many-body effects, as accounted by the GW approximation, turn the KS metal into a small gap insulator.
As this happens also in other systems, e.g. germanium, it is not a novelty in itself. 
However, here the \textit{important modification of the band shape} shown in Fig.~\ref{bandplot}\textit{d} is quite unusual:
The top-of-valence is moved away from $\Gamma$ in the plane perpendicular to $k_z$ towards a circle of points around $\Gamma$.
Similar reshaping also appears for the lowest two empty bands at $\Gamma$ and at A, and, less pronounced, for two lower bands at A and M.
Such mexican hat features have been predicted by Kohn \cite{Kohn} to occur in excitonic insulators.
Up to our knowledge, our results are the \textit{first observation of this characteristic shape from ab initio GW calculations}, pointing to an excitonic origin of the  insulating nature of TiSe$_2$.

Our (Fig.~\ref{bandplot}\textit{b}) and previous ARPES experiments \cite{Kidd,Qian} show a weakening of the signal in the neighbourhood of $\Gamma$, so that the band structure cannot be clearly resolved there.
However Kidd \textit{et al.} first remarked a loss of parabolic shape of the Se $4p$ band, together with a flattening (see Fig.~1\textit{b} and text in Ref.~\cite{Kidd}) occurring at $k \simeq$ 0.05/$\Gamma$K which is confirmed also by our ARPES spectra (Fig.~\ref{bandplot}\textit{b}).
Our GW bands (Fig.~\ref{bandplot}\textit{d}) present a loss of parabolic shape and a maximum, \textit{i.e.} a zero of the first derivative $d\epsilon/dk$, and so a flattening at $k \simeq$ 0.05/$\Gamma$M, perfectly compatible with Kidd \textit{et al.} and our ARPES findings. % (see also Ref.~\cite{TaNiSe}).
A plot of the second derivative $d^2I/dE^2$ of the ARPES signal (Fig.~\ref{bandplot}\textit{c}) presents further anomalies: at $\Gamma$ we can observe a peak at $\simeq - 0.5$ eV, to be interpreted as the light electron $\Gamma_2^{-}$ band; and then two peaks at $\simeq - 0.3$ and $\simeq - 0.2$ eV (red squares in Fig.~\ref{bandplot}\textit{c}), probably corresponding to a split of the $\Gamma_3^{-}$ Se 4$p$ degenerate bands at $\Gamma$.
%We verified that many-body effects on top of spin-orbit corrections can split those bands.
Since those bands top at $E \simeq -0.2$ eV and at $E \simeq -0.15$ eV (red circles in Fig.~\ref{bandplot}\textit{c}), we infer that a band bending should occur in order to achieve the lowest energies at $\Gamma$.
This points to a possible experimental confirmation of the GW mexican hat scenario.

For the rest of the band plot (Fig.~\ref{bandplot}\textit{a} and \textit{d}), experimental ARPES points agree much better with the GW bandplot than with DFT KS, both in the $\Gamma$M and in $\Gamma$A directions.
DFT presents empty Se $4p$ bands around $\Gamma$ and along $\Gamma$A, giving rise to a pocket of holes at $\Gamma$.
This is in striking contrast with ours (Fig.~\ref{bandplot}\textit{b}) and all previous ARPES results \cite{Rossnagel,Pillo,Kidd,Qian,Herve1} which found occupied Se $4p$ bands and no pocket of holes, exactly like in our GW calculation.
Although the question was long debated, the most recent ARPES experiments \cite{Rossnagel,Pillo,Kidd,Qian,Rasch} seem to indicate an indirect band gap of $\simeq 150$ meV between the Se $4p$ bands at $\Gamma$ and the ``emerging'' (above $E_F$ and just only termally occupied) Ti $3d$-derived band at L.
This scenario is in good agreement with our GW band structure and in contrast with DFT and oldest ARPES experiments which rather found a band overlap.
Residual mismatches of our GW and ARPES bands, like the split of the two Se $4p$ bands along $\Gamma$A, can be solved by including the spin-orbit coupling (SOC), as one can see by comparing DFT and DFT SOC bands in Fig.~\ref{bandplot}\textit{d}.
SOC and relativistic corrections have also important effects in band-crossing avoidings.
Finally, the mismatch in the position of the flat Ti $3d$ band at -2.5 eV is due to the lack of semicore electrons.
Inclusion of semicore Ti $3s$$3p$ electrons into the pseudopotential valence shell (violet dots in Fig.~\ref{bandplot}\textit{a}) improves the position of the $3d$ bands with respect to the experiment.% , while keeping unaffected the other bands.

\begin{figure}
\includegraphics[width=\linewidth]{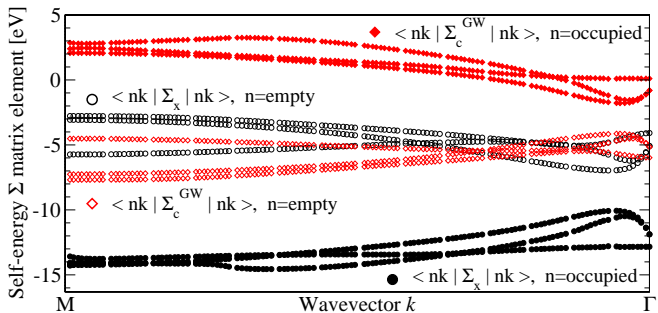}
 \caption{Exchange $\langle nk|\Sigma_x|nk\rangle$ (black circles) and correlation $\langle nk|\Sigma^\textrm{GW}_c|nk\rangle$ (red diamonds) real part contributions to the self-energy matrix elements, for the 3 topmost occupied (filled symbols) and 3 lowest empty (empty symbols) bands along M-$\Gamma$.}
 \label{sigmaxc}
\end{figure}

\paragraph{Microscopic analysis of the Mexican hat scenario}
In Fig.~\ref{sigmaxc} where we report the separate contributions of the diagonal matrix elements to the correlation $\langle nk |\Sigma_c^\textrm{GW} | nk \rangle$ and exchange $\langle nk |\Sigma_x | nk \rangle$ self-energy, entering into the GW correction to the KS energies, for the 3 topmost occupied and the 3 lowmost empty states at $\Gamma$.
It is evident that the mexican hat renormalization is dominated by exchange effects. 
Bare exchange introduce changes of 2 eV between $\Gamma$ and the new top-of-valence (TOV) circle of maxima, while screening effects included in the correlation part reduce this effect to a total value of around 0.3 eV.
Similar conclusions hold also for the two lowmost empty states at $\Gamma$, although with a softer renormalization.
If the mexican hat many-body renormalization that we see in TiSe$_2$ is really associated to its excitonic insulator nature, then it is already captured at a low level of approximation on the self-energy and there is no need to improve on correlation.
In general, one expects that vertex corrections beyond the GW approximation are required to properly describe excitons, e.g. within the Bethe-Salpeter equation approach \cite{bse} to describe the spectrum of \textit{neutral} excitations as in optical absorption.
However, in contrast to optical absorption, ARPES measures \textit{charged} excitations, and it seems that a lower level approximation, as the GW used here, can describe excitonic effects in ARPES spectra.
A Feynman diagram expansion of the GW Green's function would show that hole (electron) channels are taken into account by the GW renormalization of the electron (hole) propagator.

In order to understand the mechanism and to identify the states responsible of the mexican hat effect, we studied the matrix elements of the electric dipole operator $\hat{D} = e^{-iqr}$, directly entering into the exchange operator:
\begin{equation}
  \Sigma^x_{nk}
  = -\frac{4\pi}{V_\mathrm{BZ}}
  \sum_{n'k'} f_{n'k'} \frac{| \langle \psi_{n'k'} | e^{-i(k-k')r} | \psi_{nk} \rangle |^2}{|k'-k|^2}
, \label{exchange}
\end{equation}
with $f_{nk}$ Fermi-Dirac occupation numbers.
We studied in particular the difference of matrix elements calculated at $k=\Gamma$ and those at $k= \textrm{TOV}$ (the top of the mexican hat), as a function of $n'$ and $k'$.
We found that the mexican hat feature for the topmost $n =$ Se $4p_{3/2}$ band is mostly conjured by exchange with states on bands $n'=$ Se $4p_{3/2}$ and $4p_{1/2}$, and at $k'$ close to the Brillouin zone center and rather in the $k_xk_y$ plane (we checked the $\Gamma$M direction).
On the other hand we have found negligible the contribution coming from exchange with states at $k'$ along the $k_z$ direction ($\Gamma$A direction), and also from states faraway from $k \simeq \Gamma$ where the mexican hat feature occurs --- the Brillouin zone boundary for instance ---.
The latter are suppressed by the $|k'-k|^2$ factor at the denominator in Eq.~(\ref{exchange}).
Our analysis seems also to exclude that the mexican hat is an artifact of our $G_0W_0$ non self-consistent calculation.
Indeed the Ti $3d$-derived band, which is partially filled in DFT and becomes empty in GW, plays no role in the renormalization.
In the next iterations the increased population of the Se $4p$ bands, mostly responsible of the renormalization, and the opening of a bandgap, resulting in a reduction of the electron-hole screening, should lightly further strengthen the renormalization, or at most not change the picture. 
Overall, our analysis qualitatively confirms the physics emerging from the model calculation of Ref.~\cite{Jerome} up to the possibility of a CDW.

Finally, our calculation shows that the mexican hat effect is already present for the higher symmetry non-distorted atomic structure and/or without invoking the role of ionic degrees of freedom such as those involved in the electron-phonon interaction.
If the mexican hat is directly associated to the excitonic insulator nature of the system, it will indicate
 an instability of the ground state of purely electronic origin, without the necessity of a phonon-driven ionic mechanism.
Hence, the experimentally observed  2x2x2 distorted reconstruction \cite{DiSalvo} should rather be regarded as a consequence, following the excitons condensation and the formation of a CDW, rather than a cause.

\paragraph{Conclusions}
We presented both theoretical GW and new experimental ARPES and STS results on TiSe$_2$.
The many-body GW self-energy opens a band gap of $\approx 0.2$ eV, and strongly renormalize the band shape with the top-of-valence moved toward a circle of points away from $\Gamma$, arising in a \textit{mexican hat} feature typical of an excitonic insulator.
The calculations are in good agreement with our ARPES and STS data, which are compatible with the mexican hat scenario, as well as EELS experiments.

\paragraph{Acknowledgements}
We thank M. Gatti for useful discussions.
PA acknowledges support by the Fonds National Suisse pour la Recherche Scientifique through Div. II and MaNEP.
The high-resolution ARPES measurements were performed at the Surface and Interface Spectroscopy Beamline at the Swiss Light Source, Paul Scherrer Institute, Switzerland. 
We acknowledge Cineca for computer time and ETSF for support.


\begin{thebibliography}{22}
\bibitem{calculation-details}
We used the \texttt{ABINIT} code to perform DFT-LDA and GW pseudopotential calculations on plane waves basis.
We used Martins-Trouiller pseudopotential with an energy cutoff of 40 Ha on a 14$\times$14$\times$14 $k$-point grid for the DFT calculations.
For the GW calculation we used a cutoff of 32.8 Ha for wave functions and 2.5 Ha for dielectric and self-energy matrix elements, and we included 80, 160 bands into screening and self-energy respectively.
\end{thebibliography}
\end{document}